\documentclass{webofc}

\usepackage[varg]{txfonts}   
\usepackage{hyperref}
\usepackage[draft]{minted}
\usepackage{xcolor}
\usepackage{url}
\usepackage{changepage}
\hypersetup{colorlinks=true,citecolor=blue,urlcolor=blue,linkcolor=blue}
\begin{document}
\title{Parsing TTree Formula in Python}
%
%

\author{\firstname{Aryan} \lastname{Roy}\inst{1}\thanks{\email{aryanroy5678@gmail.com}} \and
        \firstname{Jim} \lastname{Pivarski}\inst{2}\fnsep\thanks{\email{pivarski@princeton.edu}}}

\institute{Manipal Institute of Technology
\and
          Princeton University
          }

\abstract{Uproot can read ROOT files directly in pure Python but cannot (yet) compute expressions in ROOT’s TTreeFormula expression language. Despite its popularity, this language has only one implementation and no formal specification. In a package called “formulate,” we defined the language’s syntax in standard BNF and parse it with Lark, a fast and modern parsing toolkit in Python. With formulate, users can now convert ROOT TTreeFormula expressions into NumExpr and Awkward Array manipulations.
In this contribution, we describe BNF notation and the Look Ahead Left to Right (LALR) parsing algorithm, which scales linearly with expression length. We also present the challenges with interpreting TTreeFormula expressions as a functional language; some function-like forms can’t be expressed as true functions. We also describe the design of the abstract syntax tree that facilitates conversion between the three languages. The formulate package has zero package dependencies, so we are adding it as one of Uproot's dependencies so that Uproot will be able to use TTreeFormula expressions, whether they are hand-written or embedded in a ROOT file as TTree aliases.
}
\maketitle
\section{Introduction}

\label{intro}
The C++ ROOT \cite{brun1997root} implementation allows users to store aliases along with raw data in ROOT files. These aliases behave as branches themselves, the value of which can be computed using the expression provided inside the ROOT file. These expressions are written using a unique mini-language called TTreeFormula, which is heavily inspired by C++. For example, a user can use this alias to compute the value of pt, which is defined as :

\begin{equation}
    p_t = \sqrt{p_x^2 + p_y^2}
\end{equation}

Now this expression can be written in TTreeFormula notation in multiple ways, one of them being:

{\small
\begin{adjustwidth}{4cm}{2cm}
\begin{minted}{text}
TMath::Sqrt(px*px + py*py) = pt
\end{minted}
\end{adjustwidth}
}
Uproot is a library for reading and writing ROOT files in pure Python and Numpy. When Uproot encounters these TTreeFormula expressions, it might fail because it evaluates these aliases using the Python parser (The C++ scope resolution operator would throw an error immediately). To solve this problem, a library called Formulate that can parse the ROOT mini-language was developed and with good integration, it could potentially solve Uproot's problems.

The Formulate library was built using Pyparsing \cite{pyparsing} a Python library that allows users to create and execute simple grammars by putting together a variety of classes to parse any given language. While this worked well for small expressions, the design of the original formulate was ill-suited for parsing large expressions in reasonable time. Hence, we decided to rewrite the whole library using Lark \cite{lark}, a fast and modern parsing toolkit in Python that let's us define the language syntax in standard Extended Backus–Naur form (EBNF) notation. This allowed us to use the highly performant parsers that come with Lark, greatly reducing the time taken to parse and convert ROOT TTreeFormula notation into any other format. Since the original Formulate can also parse and emit NumExpr we decided to keep it in the new implementation, so while the new Formulate can only parse two of the languages (ROOT TTreeFormula and NumExpr), it can emit all of them.

\section{Transpiling Multiple Languages}

\begin{figure}
\centering
\sidecaption
\includegraphics[width=13cm,clip]{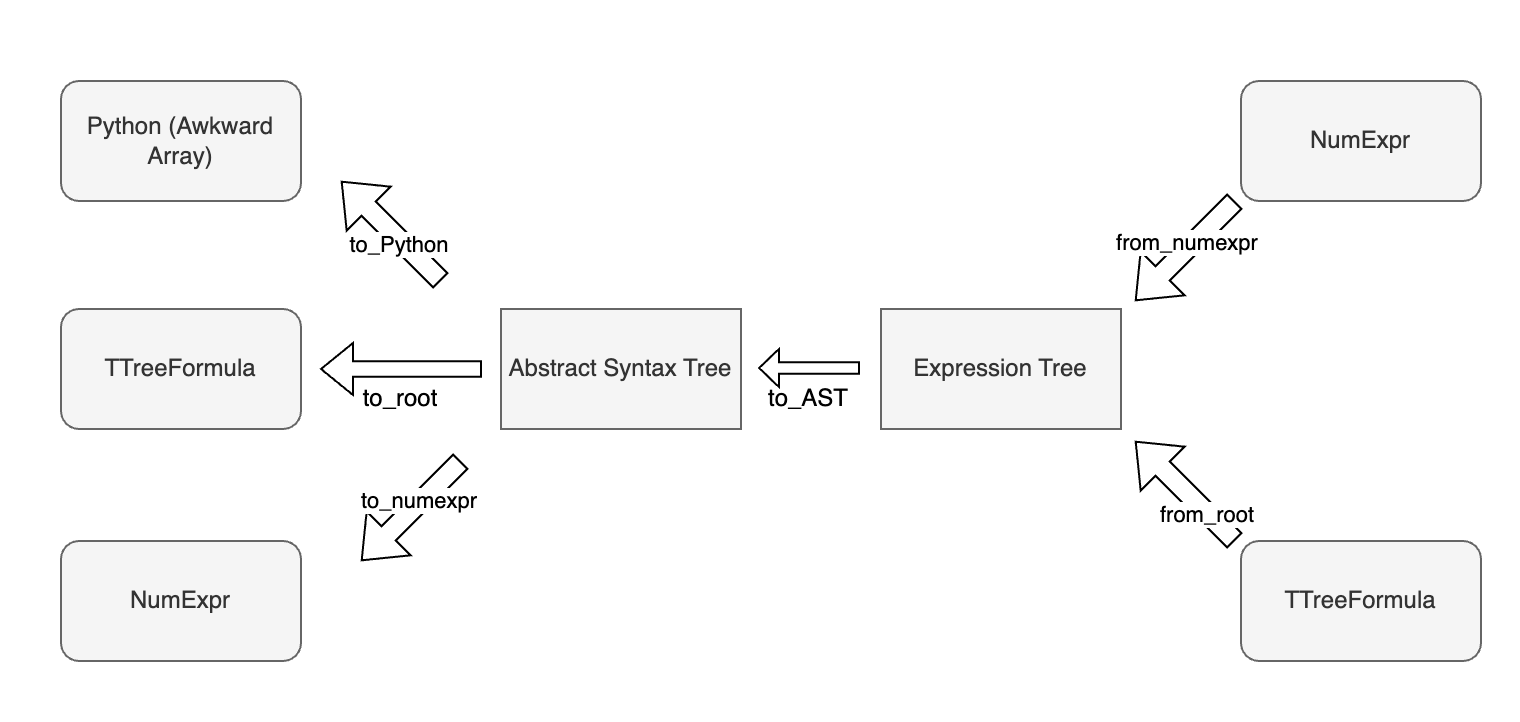}
\caption{The flow for transpiling the three languages is illustrated in this diagram. The library can parse either TTreeFormula or NumExpr using the from\_root or the from\_numexpr methods. Once parsed, an expression tree is generated by the Lark parser, which is converted into an Abstract Syntax Tree by calling the to\_AST method internally. Once we have the AST and it's nested nodes, the user can call either the to\_root, to\_numexpr or to\_python to generate strings of the translated expression.}
\label{fig-fchart}       
\end{figure}

The new Formulate has a simple design that allows for easy conversion between all the supported languages. It has a parser, that defines the formal EBNF notation for the two languages that it parses, after the expression has been parsed and an expression tree generated, we parse the expression tree to generate an Abstract Syntax Tree (AST), a data structure that can emit expressions in any of the three languages. Once the AST has been formed, the actual conversion is a matter of calling the 'to\_<language>' method on the AST Class instance as shown in Figure \ref{fig-fchart}.

\subsection{The Parser}

The parser, written using Lark, is the main point of difference between the old formulate and the new one. We have only defined the formal EBNF notation for TTreeFormula and NumExpr since we will only be parsing these two languages.

Lark implements two parsers, Earley \cite{Earley1970AnEC} and LALR(1) \cite{deremer1969practical}  that can be used to parse the EBNF grammar of our choice. While Earley is capable of parsing a wider variety of grammars, LALR was a better choice for us for the following reasons:

\begin{itemize}
    \item We do not need to define an ambiguous or non-deterministic grammar to parse either TTreeFormula or NumExpr.
    \item LALR(1) parser is very efficient in space and performance - \textit{O(n)}.
\end{itemize}

Using the LALR(1) implementation of Lark allows us to make the new Formulate a performative pure-Python package with zero dependencies. Lark allows us to generate a stand-alone LALR(1) parser directly from the grammar. This pure-Python standalone parser can then be embedded into the new Formulate without adding Lark as a dependency.

\subsection{The AST}

Once the parser has generated the expression tree, we generate the AST using an internally defined mapping from parsing tree node to AST nodes. We generate an equivalent tree using the custom AST classes that we have defined in the code base. As a rule, each unique expression generates a unique AST.

We have defined eight AST classes that act as nodes in the final AST. The individual classes correspond to eight different type of tokens that we expect to encounter while parsing the two languages.

The AST was designed to act as an Intermediate Representation (IR) for the three languages. Once generated, the AST nodes contain the language specific information to generate the correct code for any one of the supported languages. Our AST is a High Level - IR (H-IR) that holds the operands/arguments for operators/functions as data members of a Python class. This AST contains a top level root node and when the user wants code emitted in a target language, they can invoke the appropriate function which then invokes the same function in the child node which then invokes the same function in a child node ... and so on. The final code is assembled as a string which is stitched together from the output of all of these recursive function calls.
\section{Writing the Grammar}

\begin{figure}
\centering
\sidecaption
\includegraphics[width=13cm,clip]{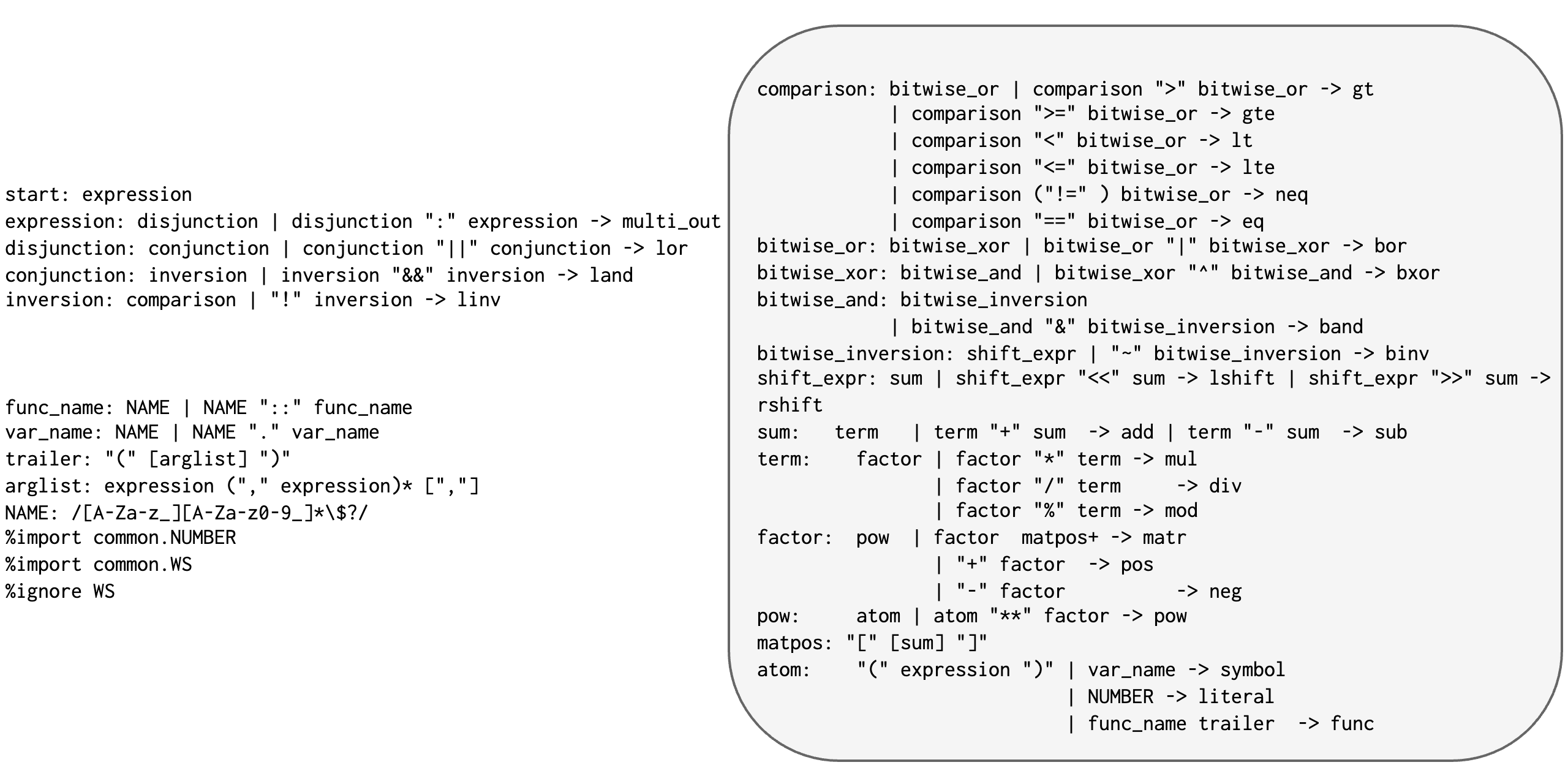}
\caption{The EBNF for both TTreeFormula and NumExpr. The parts in Grey denote the grammar for parsing NumExpr while the whole text defines the grammar for TTreeFormula. Please note that the NumExpr grammar defines func\_name as CNAME, after which the NumExpr grammar terminates.}
\label{fig-ebnf}       
\end{figure}

Parsing TTreeForumla and NumExpr syntax required writing the EBNF that covered all the possible syntactical combinations that either language could present. The operator precedence in both the languages is similar to what the Python grammar expects, hence we took the Python BNF as reference for both the languages.

NumExpr's syntax is modeled after the expression evaluation in Python. TTreeFormula on the other hand is a superset of NumExpr's allowed tokens as shown in Figure \ref{fig-ebnf}.  Hence, first we will talk about the NumExpr grammar and then move on to TTreeFormula.

The NumExpr grammar gives the highest precedence to the bitwise comparison operators. All of them have equal precedence followed by the shift operator then addition and subtraction operators. NumExpr allows functions of only one form, i.e "function\_name(argument\_name)". While we could have used the TTreeFormula grammar to parse NumExpr expressions as well, to felicitate better error messages and to avoid any problems in case we discover some differences in how either of the languages work in the future, we decided to to go with separate parsers for both of them.

While the TTreeFormula mini-language has been codified before, there is no formal grammar for it. Therefore to write a standard EBNF for it, we had to carefully assess the behavior of each operator before we could translate it to any other language. In the grammar, the highest precedence is given to the "multiple output" operator. This was TTreeFormula specific and made sense to be given the highest priority since it can contain any number of general expression with that many outputs.

The TTreeFormula grammar also contains the logical comparison operators, something that NumExpr lacks. While the middle portion of the grammar is the same, we see some more differences towards the end where the function forms are defined. In addition to the general function form also found in the NumExpr grammar, TTreeFormula allows another kind of function form; "token1::token2(token3)", containing the scope resolution operator from C++. This was necessary to parse TTreeFormula, which as a language was heavily inspired by C++.

\section{Performance}

\begin{figure}
\centering
\sidecaption
\includegraphics[width=13cm,clip]{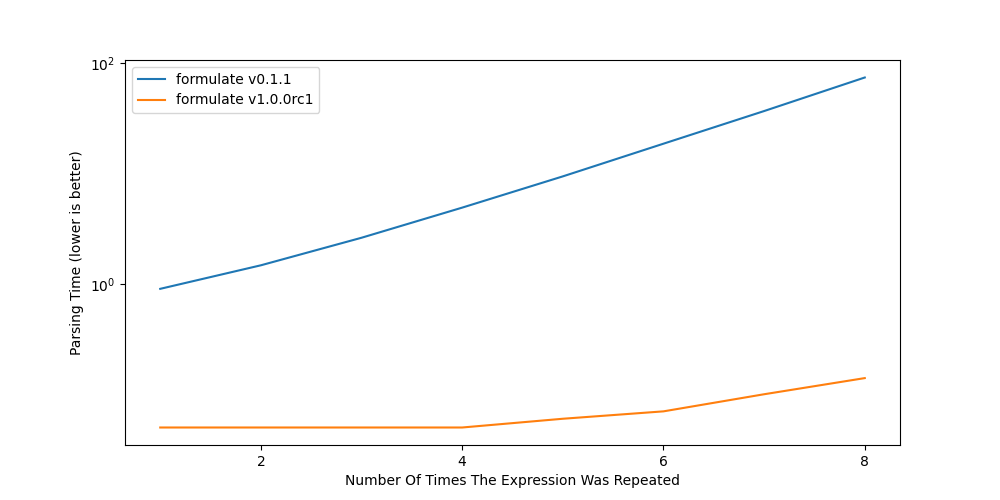}
\caption{The previous implementation of Formulate scales exponentially with the number of tokens in the parsing expression whereas the curve for the new implementation is much less aggressive. The y-axis is on a log scale. The x-axis indicates how many times the original expression (54 characters) has been repeated}
\label{fig-perf}       
\end{figure}

Performance of the old Formulate was the biggest driving force behind the development of the new Lark-based Formulate. The new Formulate uses LALR(1), which is one of the least algorithmically complex ways to parse the kind of grammar we have written for Formulate. As is clearly visible in the Figure \ref{fig-perf}, the switch to LALR(1) shows a remarkable speedup.

As mentioned before, the Lark parser uses LALR(1), an algorithm that has the time complexity of \textit{O(n)} whereas Pyparsing used the Parsing Expression Grammar (PEG) \cite{10.1145/964001.964011} parser that can exhibit exponential time complexity as shown in the plot. The graph indicates a much lower start-up time for Lark compared to Pyparsing. The stark difference in the slope of both the curves indicates the much better scaling capability of the Lark-based Formulate.

The expressions for this comparison are inspired from real life use cases of formulate. The comparison was done by adding the same base expression to itself to increase the length of the input string and clocking the time taken by both the versions of formulate to parse the expressions. The base-expression used was:

{\small
\begin{adjustwidth}{2.25cm}{2cm}
\begin{minted}{text}
((weight * (n_mu > 0)) * ((tt_cat + tt_cat + tt_cat)))
\end{minted}
\end{adjustwidth}
}

\section{Conclusion}

In this contribution, we described the implementation details and challenges of incorporating Lark, a fast and efficient pure python parsing library into Formulate. The new Formulate was built to overcome the severe performance issues we faced with the old design based on an implicit grammar definition.
It should also be noted that this is the first time that the grammar for TTreeFormula has been defined in EBNF. The new and improved Formulate can handle complex expressions and shows an average speedup of up to 526 fold when tested on large expressions inspired from real world examples.

\section{Acknowledgment}
This work was supported by the National Science Foundation under Cooperative Agreement
PHY-2323298 (IRIS-HEP).

%
%
%
\bibliographystyle{plain} 
\bibliography{refs} 

\begin{thebibliography}{6}

\bibitem{brun1997root}
R.~Brun, F.~Rademakers, Root—an object oriented data analysis framework, Nuclear instruments and methods in physics research section A: accelerators, spectrometers, detectors and associated equipment \textbf{389}, 81 (1997).

\bibitem{pyparsing}
P.~McGuire, pyparsing: A python parsing module (2025), accessed: 2025-02-16, \urlstyle{tt}\url{https://github.com/pyparsing/pyparsing}

\bibitem{lark}
E.~Shinan, Lark: A parsing toolkit for python (2025), accessed: 2025-02-16, \urlstyle{tt}\url{https://github.com/lark-parser/lark}

\bibitem{Earley1970AnEC}
J.~Earley, An efficient context-free parsing algorithm, Communications of the ACM \textbf{13}, 94  (1970).

\bibitem{deremer1969practical}
F.L. DeRemer, Ph.D. thesis, Massachusetts Institute of Technology (1969)

\bibitem{10.1145/964001.964011}
B.~Ford, Parsing expression grammars: a recognition-based syntactic foundation, in \emph{Proceedings of the 31st ACM SIGPLAN-SIGACT Symposium on Principles of Programming Languages} (Association for Computing Machinery, New York, NY, USA, 2004), POPL '04, p. 111–122, ISBN 158113729X, \urlstyle{tt}\url{https://doi.org/10.1145/964001.964011}

\end{thebibliography}

\end{document}